\documentclass[prb, aps, twocolumn, amsmath, amssymb, superscriptaddress,noeprint]{revtex4-2}

\usepackage{float}
\usepackage[dvips,final]{graphicx}
\usepackage{dcolumn}% Align table columns on the decimal point
\usepackage{bm}% bold math
\usepackage{textcomp}
\usepackage{amsmath,amssymb,amsfonts}
\usepackage{color}
\usepackage{hyperref} % add hypertext capabilities
\hypersetup{colorlinks,linkcolor={blue},citecolor={blue},urlcolor={blue}}  

\newcommand{\OO}[1]{\mathcal{O}\left(#1\right)}

\begin{document}

\title{Chirped amplitude mode in photo-excited superconductors}

\author{Thomas Blommel}
\affiliation{Department of Physics, University of Michigan, Michigan, USA}

\author{Jason Kaye}
\affiliation{Center for Computational Quantum Physics, Flatiron Institute, 162 5th Avenue, New York, NY 10010, USA}
\affiliation{Center for Computational Mathematics, Flatiron Institute, 162 5th Avenue, New York, NY 10010, USA}

\author{Yuta Murakami}
\affiliation{Center for Emergent Matter Science, RIKEN, Wako, Saitama 351-0198, Japan}

\author{Emanuel Gull}
\affiliation{Department of Physics, University of Michigan, Michigan, USA}

\author{Denis Gole{\v z}}
\affiliation{Jo{\v z}ef Stefan Institute, SI-1000, Ljubljana, Slovenia}
\affiliation{Faculty of Mathematics and Physics, University of Ljubljana, 1000
Ljubljana, Slovenia}

\date{\today}%
\begin{abstract}
Using a state-of-art numerical scheme, we show that the Higgs mode under excitation exhibits chirped oscillations and exponential decay when fluctuations are included.  This is in stark contrast to conventional BCS collisionless dynamics which predict power-law decay and the absence of the chirping. The chirped amplitude mode enables us to determine the local modification of the effective potential even when the system is in a long-lived pre-thermal state. We then show that this chirped amplitude mode is an experimentally observable quantity since the photo-induced (super)-current in pump-probe experiments serves as an efficient proxy for the order parameter dynamics, including the chirped dynamics. Our result is based on the attractive Hubbard model using dynamical mean-field theory within the symmetry-broken state after a excitation across the superconducting gap. Since the collective response involves long timescales, we extend the hierarchical low-rank compression method for nonequilibrium Green’s functions to symmetry-broken states and show that it serves as an efficient representation despite long-lived memory kernels.
\end{abstract}

\maketitle

\section{ Introduction} Soon after the seminal paper on BCS superconductivity~\cite{bardeen1957}, Anderson pointed out that the electromagnetic response of a superconductor leads to a collective  response~\cite{anderson1958,anderson1958b,anderson1963}.  This realization had a profound impact on condensed matter physics, as well as particle physics, with the prediction of the Higgs mode~\cite{Higgs1964,higgs1964b}. 
It took almost two decades before such an excitation was observed in 2H-NbSe$_2$ with the coexistence of superconductor and charge-density-wave~\cite{sooryakumar1980,Littlewood1981,Littlewood1982,pekker2015} phases, and another three decades for direct observation using terahertz spectroscopy either in a pump-probe setup~\cite{matsunaga2013} or with third-harmonics generation~\cite{matsunaga2014,matsunaga2017}. 

This progress opened the field of Higgs spectroscopy, which has now been applied to numerous superconducting materials, including superconductors with running supercurrents~\cite{moor2017,nakamura2019,Crowley2022} and unconventional superconductors~\cite{katsumi2018,chu2020, Poniatowski2022}. Ever more precise measurements of collective mode response enable the study of their long-time behavior after excitation both in solids~\cite{matsunaga2013} as well as quantum simulators~\cite{Young2024,lewis2021,behrle2018}. 

The theoretical description of the time evolution of the Higgs mode is difficult.  Time-dependent Landau-Ginzburg theory assumes an ad-hoc rapid modification of the effective potential and damping is considered on the phenomenological level~\cite{aranson2002,zong2019a}. Microscopic time-dependent BCS theory predicts richer dynamics, including a power-law decay~\cite{volkov73,tsuji2015} or solitonic solution~\cite{barankov2004,yuzbashyan2006,barankov2006,yuzbashyan2015}. However, all these theoretical statements assume dissipationless electronic dynamics, which is questionable in real materials and opens up a question on the decay of collective modes upon the inclusion of electronic scattering.

The main difficulty of the theoretical description of electronic scattering on collective mode dynamics stems from the fact that collective response involves long timescales. This has limited previous studies to dissipationless dynamics~\cite{volkov73,tsuji2015,barankov2004,yuzbashyan2006,barankov2006}, or when fluctuations were included, the description was limited to relatively short times with few coherent oscillations and limited information about amplitude mode lifetime~\cite{Kemper2017,nosarzewski2017,kumar2019,Murakami2016-2}.

Only recently, progress in numerical solvers for nonequilibrium Green's functions~\cite{Nessi}, including the generalized Kadanoff-Baym ansatz~\cite{Lipavski1986,tuovinen2020,Schluenzen2017,pavlyukh2022} and memory kernel truncation methods~\cite{Schueler2018,stahl2022memory},
has allowed access to long enough times to examine these questions. However, the application of these techniques to symmetry-broken phases is much less understood due to the long-lived power-law correlators acting as memory kernels.

In this work, we study the long-time evolution of collective modes by solving the full Kadanoff-Baym equations (KBE), and show that the order parameter displays an extremely slow approach to a thermal state. Within such a nonthermal state,
the amplitude mode exhibits chirping~(dynamical decrease in the frequency). 
The velocity of the chirping increases as we excite the system with larger field strengths.
We show that the effect is experimentally accessible in terahertz pump-probe experiments by measuring the probe-induced currents, which can be used
to extract the dynamics of the order parameter with a small imprint of the quasiparticle response.

\begin{figure*}
    \centering
    \includegraphics[width=0.95\textwidth]{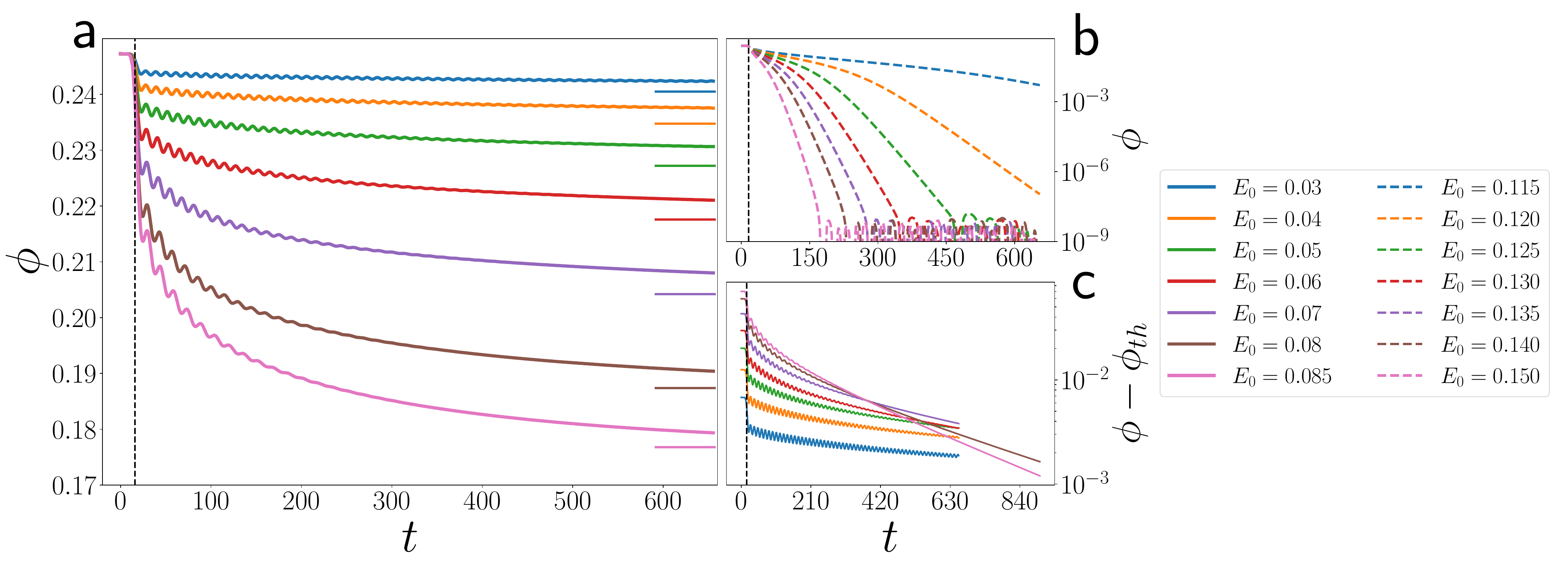}
    \vspace{-0.4cm}
    \caption{Time evolution of the order parameter $\phi$ after photoexcitation with pump amplitude $E_0$ in the (a) weak excitation regime, in which amplitude mode oscillations are present, and (b) strong excitation regime, in which we observe superexponential decay.  The noise at small values in (b) is consistent with the truncation tolerance $10^{-8}$ used in the HODLR compression scheme.    (c) The order parameter of the system exponentially approaches the expected thermal value, $\phi_{th}$, which is the value of the order parameter in the equilibrium system that has the same energy (horizontal bars on right-hand side of (a)).  $E_0 = 0.08, 0.085$ are integrated further in time to emphasize the exponential decay. }
    \label{fig:order-param}
\end{figure*}

We also highlight our use of the hierarchical off-diagonal low-rank (HODLR)
compression method for the numerical solution of the KBE~\cite{Kaye2021}.
The standard $\OO{N^3}$ scaling of the direct solution of the KBE with the number of time steps $N$, resulting from memory terms, typically prevents access to the time scales
necessary to analyze the lifetime of collective excitations~\cite{Nessi,Kemper2017,nosarzewski2017,kumar2019}. A new class of data compression-based methods, including the HODLR approach, have recently been proposed to overcome this bottleneck \cite{yin22,reeves23,Shinaoka2023,murray2023,meirinhos22}. Viewing two-time Green's functions as matrices, the HODLR
method decomposes these matrices into blocks, refined towards the diagonal,
which in many cases have been observed to be numerically low-rank. A truncated singular value
decomposition of these blocks is systematically updated on-the-fly
during time-stepping and used to compute history integrals with controllable accuracy,
yielding an $\OO{k^2 N^2 \log N}$ computational complexity and $\OO{k N \log N}$
memory complexity, for $k$ the maximum block rank, without modifying the underlying KBE. 
Using this method, we demonstrate a 100-fold decrease in the computational cost required to propagate to the time scales studied compared with direct time stepping, with 50-200 times less memory.

\section{ Model and method}\label{sec:Model}
We study superconductivity within the attractive Hubbard model 
\begin{equation}\label{Eq:Model}
     H = -t_0 \sum_{\langle j,k\rangle\sigma}e^{iqA(t)} d^\dagger_{j\sigma}d_{k\sigma}-U\sum_j (n_{j\uparrow}-1/2)(n_{j\downarrow}-1/2),
\end{equation}
where $d_{i\sigma}$ is the annihilation operator at site $i$ and spin $\sigma$,
$n_{i\sigma}$ is the spin-dependent density operator and $q$ is the charge. The energy scale is set via the hopping amplitude $t_0=1$, the Coulomb attraction is set to $U=2$,
and we  fix the occupation at half-filling. Units are measured in $\hbar=q=1.$

We solve the problem within time-dependent dynamical mean-field theory~(DMFT) on
the Bethe lattice with infinite coordination number, using the Nambu formalism to describe the s-wave superconducting phase.
This leads to matrix-valued Green's functions
$G_{\alpha\beta}(t,t')=-i\langle\mathcal{T} \psi_{\alpha}(t)
\psi^{\dagger}_{\beta}(t') \rangle,$ with the Nambu spinor
$\psi=\{d_{\uparrow},d^{\dagger}_{\downarrow}\}$. For the impurity solver, we
use the self-consistent second-order perturbation theory known as second
Born, or GF2~\cite{Dahlen2006,Dong2022,Rusakov2016}. Collective order parameter
dynamics take place on long time scales, and such long-time integration of the
KBE is made possible by using a compressed representation
of the Green's function to reduce both the computational
and memory complexity~\cite{Kaye2021}. We refer to the SM for further
implementation details.  

We induce a dynamical perturbation by a short pump pulse introduced by a Peierls
substitution in Eq.~\ref{Eq:Model}. The electric field $E_p=-\partial_t A$
is parameterized as 
\begin{equation}
    E_P(t) = E_0 \, \text{exp}\left(-\frac{(t-t_c)^2}{\sigma^2}\right)\text{sin}(\omega (t-t_c)),
    \label{Eq:Epump}
\end{equation}
with pulse center $t_c$, pulse width $\sigma$, driving frequency $\omega$, and amplitude $E_0$.

\begin{figure*}
    \centering
    \includegraphics[width=0.95\textwidth]{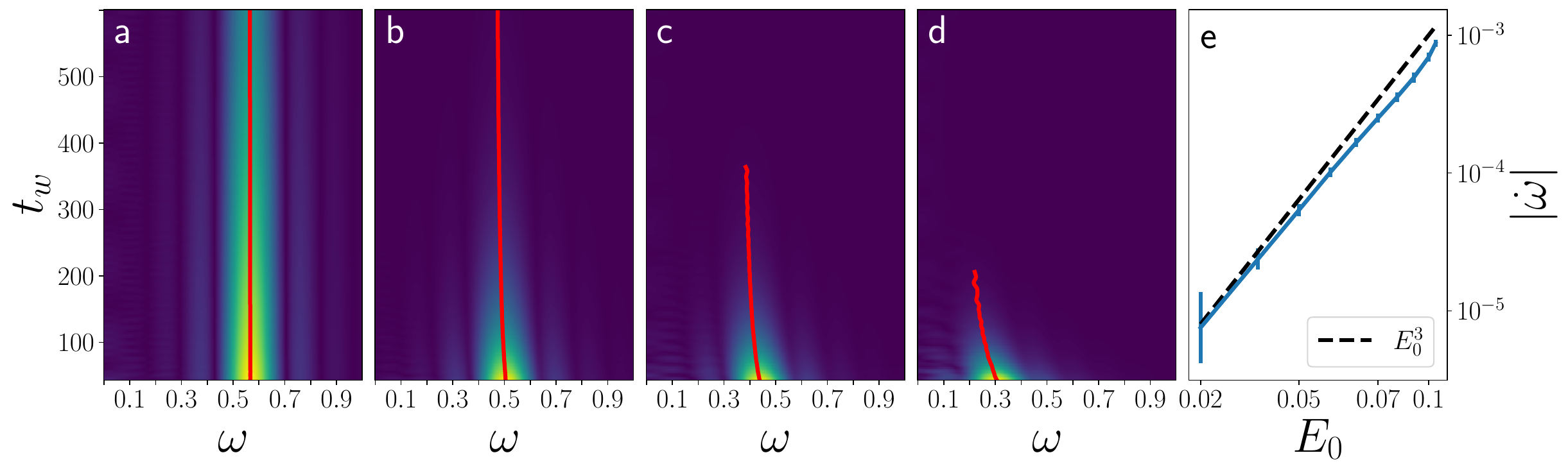}
    \caption{(a-d): Windowed Fourier transform $F[\tilde{\phi}](t_w)$ of the order parameter centered around $t_w$ with exponential background subtracted, for pump amplitudes $E_0=0.02$, $0.06$, $0.08$, and $0.105$. The red line tracks the maximum at each value of $t_w$. (e) Initial chirping velocity, defined as the slope of the red lines in (a-d) for a small window of time immediately after the pump pulse.  Error bars are obtained by varying width of FT window and number of time-points used for the linear fit.}
    \label{fig:chirp}
\end{figure*}

\section{Results} For an interaction strength of $U=2$ the attractive Hubbard model in equilibrium has a phase transition at inverse temperature $1/T_c = \beta_c\approx10.6$ within the GF2 approximation.  Across this transition, the system develops a nonzero superconducting pair amplitude $\phi(t)=\sum_k \langle c_{k\downarrow} c_{-k\uparrow}\rangle$, which from now on we consider as the order parameter.  We choose a temperature $\beta=18$ deep in the ordered state, and find that the system has a band gap of $2\Delta\approx 0.65$.

To excite the amplitude mode, we pump the system with a pulse which has a frequency twice the band gap~($t_c^{\mathrm{Pump}}=16$, $\sigma=6.5$, $\omega=4\Delta$). The time evolution of the order parameter is presented in Fig.~\ref{fig:order-param}.  At weak excitation strength, the order parameter decays at an extremely slow exponential rate~\cite{tsuji2013,Werner2012afm}. The expected thermal values $\phi_{th}$, marked by the horizontal lines, are not reached on the time scale of our simulation; see also Fig.~\ref{fig:order-param}(c), in which the thermal values are subtracted.
For intermediate pump strengths, we observe the decay of the order parameter through several orders of magnitude.  On top of the exponential relaxation, well-defined oscillations correspond to the amplitude mode excitation~\cite{shimano2020,pekker2015}.
As the pump strength is increased, the initial amplitude of these oscillations increases; however, they also have shorter lifetimes. Increasing the pump amplitude further, we reach a point at which the lifetimes of these oscillations hinder our ability to study their frequency~\cite{tsuji2013, golez2016,Picano2021b}, see Fig.~\ref{fig:envelope}(a,b) in the SM.

We observe that the frequency of the collective mode oscillations
gradually chirps to smaller values. To extract the chirping velocity, we first subtract the background exponential decay from the data in Fig.~\ref{fig:order-param}, denoted as $\tilde{\phi}$, and then compute a windowed Fourier transform $\text{FT}[\tilde{\phi}](t_w)=\int_0^{T_{max}} dt \, \tilde{\phi}(t) e^{- i\omega t- (t-t_w)^6/\sigma^6_w}$.  The width of the window, $\sigma_w=24$, is set such that at least 4 oscillations of the order parameter are captured at any time, $t_w$.  We define the chirping velocity $\dot{\omega}$ as the slope of the maximum of $\text{FT}[\tilde{\phi}](t_w)$, which is tracked in Fig.~\ref{fig:chirp}(a-d) for increasing pump amplitudes.  Fig.~\ref{fig:chirp}(e) shows that the chirping velocity follows a power-law scaling $\dot \omega\propto E_0^{3}$ with the pump amplitude, which is the same as the scaling $j\propto E_0^3$ of the current due to photo-induced amplitude mode oscillations~\cite{tsuji2015,shimano2020}.

Whereas the dissipationless BCS dynamics lead to either power-law decay~\cite{volkov73,gurarie2009} or persistent oscillations~\cite{barankov2004,barankov2006,yuzbashyan2015,yuzbashyan2006} of the superconducting amplitude depending on the initial condition, the inclusion of fluctuations modifies the Higgs mode dynamics into an exponential decay, see SM.
This is consistent with the effect of thermal noise on the dissipationless BCS dynamics~\cite{barankov2004,Murakami2016-3}. The lifetime of the amplitude mode decays with increasing pump strength (see Fig.~\ref{fig:chirp}(a-d)), and changes by more than an order of magnitude from the weak to the strong excitation regime.  

The chirping of the amplitude mode is the main result of this work. Observing such phenomena requires microscopic treatment of fluctuations beyond the mean-field theory, see SM for comparison, as the resulting state is highly nonthermal. Moreover, the chirping phenomenon is not limited to superconductors but should be observed in many systems with broken symmetry. For example, the superconducting solution analyzed in this work can be exactly mapped to the evolution of the antiferromagnetic state in the repulsive Hubbard model~\cite{Shiba1972}, and it is degenerated with the charge density waves. Both of these phases have been addressed experimentally ~\cite{hong2017,demsar1999,yusupov2010,yoshikawa2021}.
In actual experimental setups, there is a trade-off in pump strength between the chirping velocity and the lifetime of the amplitude mode, which depends on the details of the pulse and the material studied. 

Now, we interpret the microscopic dynamics within the time-dependent Landau-Ginzburg theory. The chirping of the amplitude mode 
corresponds to a dynamical reduction of the local curvature due to the feedback of the electronic scattering on the collective order. While it would be tempting to interpret the chirping as a heating effect within the Landau-Ginzburg theory, we emphasize that the system is closed and in a prethermal regime, 
which the temperature change cannot describe. In principle, one could construct a time-dependent Landau-Ginzburg theory of chirping. However, it is difficult in practice to extract the evolution of global structure of the Landau-Ginzburg potential, even when microscopic dynamics are available.
An important consequence of our analysis is that the chirped frequency can act as a microscopic measure of the dynamically modified local curvature.

\begin{figure}
    \centering
    \includegraphics[width=0.5\textwidth]{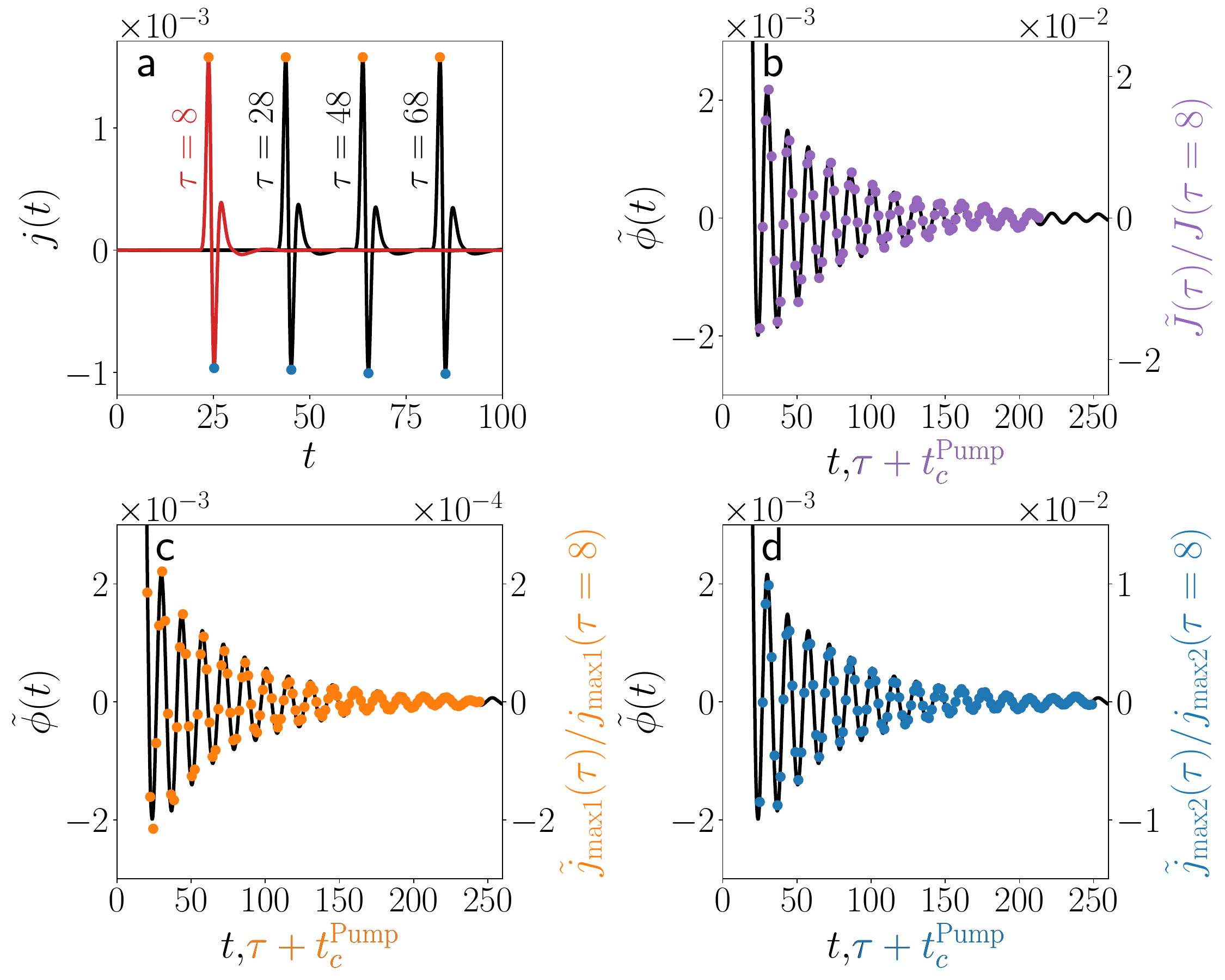}
    \caption{(a) Current, $j$, induced by a monocycle probe pulse with parameters $E_0=0.1E^{\text{Pump}}_0$, $t_c = t^{\text{Pump}}_c+\tau$, $\sigma = 1$, $\omega=1$, $E^{\text{Pump}}_0=0.08$.  The markers denote the location of the maximal current response.
    (b) Integrated current response with background subtracted off, $\tilde{J}$, compared to the order parameter oscillations, $\tilde{\phi}$. (c) Oscillations of the first maximum of the current response to the probe field, $j_{\mathrm{max}1}$, corresponding to the orange points in (a).  (d) Oscillations of the second maximum of the current response to the probe field, $j_{\mathrm{max}2}$, corresponding to the blue points in (a).}
    \label{fig:pp-exp}
\end{figure}

We next discuss how this physics could be observed experimentally. The amplitude mode does not couple linearly to light; however, it can be observed in nonlinear processes, such as pump-probe setups~\cite{matsunaga2013} or the third-harmonics response~\cite{matsunaga2014,shimano2020,tsuji2015}. Our numerical tests agree with previous observations that it
is difficult to extract the amplitude mode signal from the optical response due to the overlap with the quasiparticle continuum~\cite{kumar2019}.  Following previous studies~\cite{matsunaga2013}, we rather focus on  the photo-induced current and show that it serves as an excellent proxy for the dynamics of the order parameter.
While, in principle, the optical conductivity and photo-induced current carry similar information, we found that in practice it is much easier to extract coherent oscillations from the time-dependent current, compared with spectral analysis of the optical conductivity.

We evaluate the probe-induced current by directly simulating an additional probe pulse at a time $\tau$ after the pump pulse, and measuring the induced current~\cite{lenarcic2014,shao2016,eckstein2008}. This approach is theoretically convenient as it includes vertex corrections to the response function, which represents the contribution from the collective modes~\cite{murakami2016,golez2020}.  We propose two measurement protocols and show that the current response in both cases could be connected with the collective dynamics of the amplitude mode.  The probe pulse is modelled as a small monocycle pulse, given by Eq.~\ref{Eq:Epump} with parameters $\sigma=1$, $t_c=t_c^{\mathrm{Pump}}+\tau$, $\omega=1$, and $E_0=0.1E^{\text{Pump}}_0$.  Fig.~\ref{fig:pp-exp}(a) shows the current response to such probe field for several different delay times $\tau$.  We track the dependence of the maximal current response $j(t_\mathrm{max})$ versus the delay time and observe decaying oscillations that align to high accuracy with the order parameter dynamics, see Figs.~\ref{fig:pp-exp}(c-d). The maximum current $j(t_\mathrm{max})$ thus gives information about both the chirped frequency and the lifetime of the amplitude mode.  We also confirmed that the time-averaged current response $J(\tau)=\hat{j}(\omega=0;\tau)$ shows excellent agreement with the dynamics of the order parameter, shown in Fig.~\ref{fig:pp-exp}(b).

Based on these results, we propose that the direct measurement of the
photo-induced current is a convenient way to measure both the chirping of the amplitude mode as well as to decay envelope of the Higgs mode which would be power-law~(exponential) in the dissipationless~(fluctuating) description.
The photo-induced current and chirping can be measured in the pump-probe setup using either free-space electro-optic sampling, like in Ref.~\cite{matsunaga2013}, or transmission line experiments, which were recently used to detect superconducting nonlinear transport~\cite{wang2023}, the light-induced anomalous Hall effect due to circularly polarized light~\cite{McIver2020}, and ultrafast resistive switching in 1T-TaS$_2$~\cite{venturini2022}.  In these experiments, one measures the variation of the transmitted field induced by the probe field in the material. The modification of the transmitted field originates from the variation of the microscopic polarization in the material which is connected with the microscopic currents\cite{Tanaka2012,Benfatto2019,Felderhof1987}.

\begin{figure}
    \centering
    \includegraphics[width=0.5\textwidth]{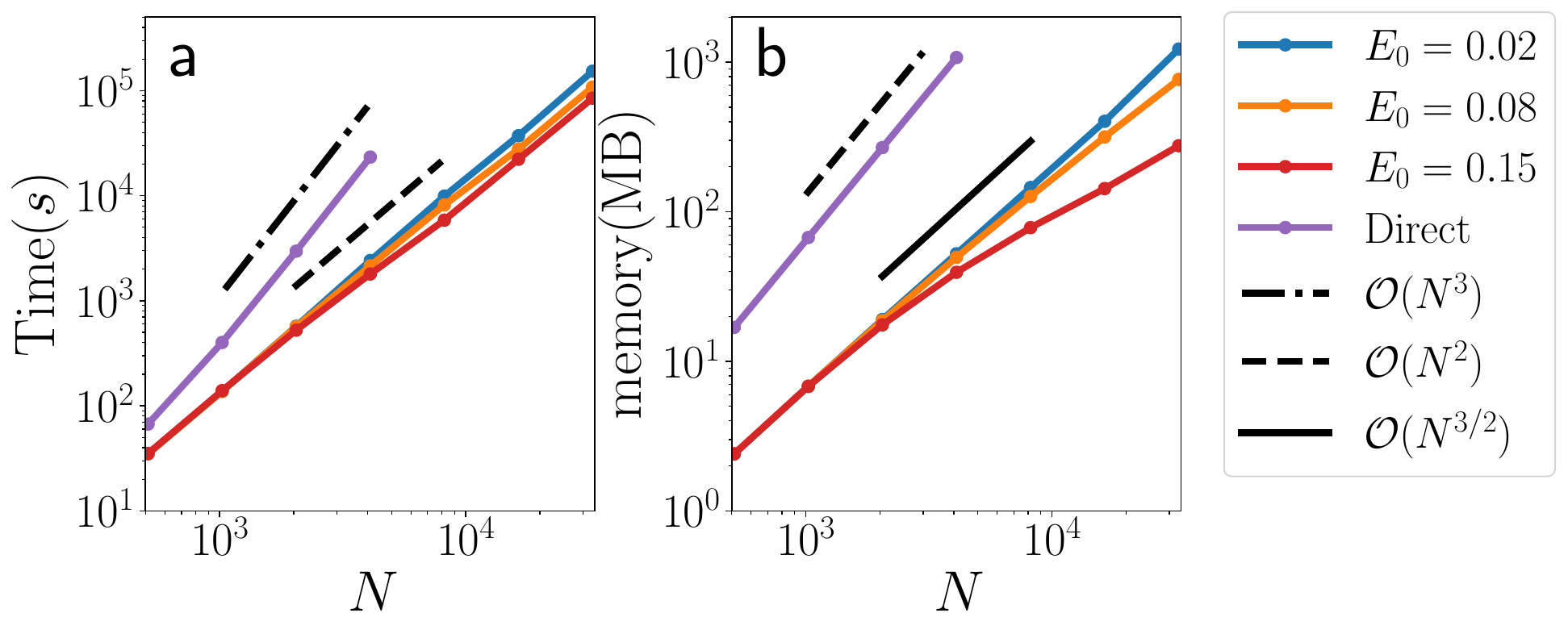}
    \caption{(a) Wall clock time and (b) required memory with increasing number of timesteps for HODLR propagation scheme versus direct time stepping, for various pump amplitudes, $E_0$, starting from the equilibrium state at $\beta=18$.}
    \label{fig:scaling}
\end{figure}

Since our analysis was enabled by the HODLR compression technique for the numerical solution of the KBE \cite{Kaye2021}, we consider the compressibility of the dynamics of symmetry-broken states, and the performance of the scheme compared with the direct solution of KBE.
In Fig.~\ref{fig:scaling}(a), we observe approximately quadratic scaling of
the computational cost with
the propagation time over a range of pump intensities and equilibrium state
temperatures. We reach 32768 time steps in roughly one day on a
single-core of a workstation using 0.5 GB of memory. By contrast, we extrapolate that
cubic-scaling direct time stepping using the NESSi code~\cite{Nessi} would have
taken 129 days with 137 GB of memory. Fig.~\ref{fig:scaling}(b) shows the different scaling between the full Green's function and the HODLR representation.  At $N=32768$ we have a factor of $50$-$200$ memory compression of the Green's function and self-energy compared with direct storage on a two-time grid.  
The SM considers the compressibility in more detail.

We have shown that the collective amplitude mode displays chirped dynamics after the above gap excitation and have analyzed a direct experimental signature of the response based on the photo-induced current within the pump-probe setup. Future applications include the nonlinear response of unconventional superconductors~\cite{katsumi2018,chu2020}, and extensions beyond dynamical mean-field theory to study spatial fluctuations~\cite{stahl2021} and the build-up of the fluctuating order in systems with competing orders~\cite{sun2020,Zong2019b}. A particularly appealing question is the instability of the system to form inhomogeneous patterns after excitation~\cite{dzero2009,chern2019,Barresi2023} that should be extremely sensitive to the decay mechanism of the Higgs mode presented in this work. Methodologically, this work represents the first application of the HODLR compression technique for the KBE to symmetry-broken states, and demonstrates that it can access long enough time scales for the practical study of a broad field of photo-induced phase transitions~\cite{Nasu2004,murakami2023,torre2021b,Giannetti2016}.

\section{Acknowledgments}
DG acknowledges the support of the projects J1-2463, N1-0318, MN-0016-106, and P1-0044 program of the Slovenian Research Agency.
TB was funded by
the Department of Energy via Grant No. DE-SC0020347 until Aug. 2023. TB as of Aug. 2023 and EG were supported by the U.S. Department of Energy, Office of Science, Office of Advanced Scientific Computing Research and Office of Basic Energy Sciences, Scientific Discovery through Advanced Computing (SciDAC) program under Award Number(s) DE-SC0022088.
YM is supported by Grant-in-Aid for Scientific Research from JSPS, KAKENHI Grant Nos. JP20K14412, JP21H05017 and JST CREST Grant No. JPMJCR1901.
The Flatiron Institute is a division of the Simons Foundation.

\appendix
\section{Implementation details}\label{App:Details}
The attractive Hubbard Hamiltonian in Eq.~(\ref{Eq:Model}) can be rewritten as 
\begin{align}
H &= -t_0 \sum_{\langle j,k\rangle\sigma}\sigma e^{iqA(t)\sigma}\psi^\dagger_{j\sigma}\psi_{k\sigma}\\&+U\sum_i \psi^{\dagger}_{i\downarrow}\psi_{i\downarrow}\psi^{\dagger}_{i\uparrow}\psi_{i\uparrow}-\frac{U}{2}\sum_{i\sigma}\psi^\dagger_{i\sigma}\psi_{i\sigma},\nonumber
\end{align}

where we have used the Nambu spinors
\begin{gather}
    \psi_i = \begin{pmatrix}
        d_{i\uparrow}\\
        d^\dagger_{i\downarrow}
    \end{pmatrix}.
\end{gather}

We again use the Nambu spinors to define the anomalous Green's function \begin{equation}
    G(t,t') = \langle\psi(t)\psi^\dagger(t')\rangle = \begin{pmatrix}
        \langle d_{\uparrow}d^\dagger_{\uparrow}\rangle(t,t') && \langle d_{\uparrow}d_{\downarrow}\rangle(t,t') \\
        \langle d^\dagger_{\downarrow}d^\dagger_{\uparrow}\rangle(t,t') && \langle d^\dagger_{\downarrow}d_{\downarrow}\rangle(t,t')
    \end{pmatrix},
\end{equation}
which is the quantity we compute by solving the KBE.

We solve this system within the DMFT approximation on a Bethe lattice, giving the hybridization function \begin{gather}
    \Delta(t,t') = \Delta_R(t,t') + \Delta_L(t,t')\\
    \Delta_R(t,t') = \frac{1}{2}\Bar{t}_0(t)\sigma_z G(t,t') \sigma_z \Bar{t}_0^*(t')\\
    \Delta_L(t,t') = \frac{1}{2}\Bar{t}_0^*(t)\sigma_z G(t,t') \sigma_z \Bar{t}_0(t'),
\end{gather} with hopping matrix elements given by
\begin{align}
    \Bar{t}_0 &= \begin{pmatrix}
        e^{iA(t)} && 0 \\
        0 && e^{-iA(t)}
    \end{pmatrix},\\
    A(t) &= -\int_0^t E(s) ds
\end{align}
due to the Peierls substitution. We set the charge and the lattice distance to unity, $q=a=1$.

The impurity problem is solved using the fully self-consistent second Born approximation. In this approximation, the Fock term is given by
\begin{equation}
    \Sigma^F_{ij}(t) = i U(t) G^<_{ij}(t,t^-) \delta_{i\Bar{j}},
\end{equation} and the Hartree term cancels the chemical potential since we are at half-filling.  There are two second-order diagrams:
\begin{align}    \Sigma^B_{ij}(t,t') &=   U(t) U(t') G_{ij}(t,t') G_{\Bar{i}\Bar{j}}(t,t') G_{ij}(t',t),\\
    \Sigma^E_{ij}(t,t') &= - U(t) U(t') G_{\Bar{i}j}(t,t') G_{i\Bar{j}}(t,t') G_{\Bar{i}\Bar{j}}(t',t).
\end{align}

The hybridization $\Delta[G](t,t')$ and the self-energy $\Sigma[G](t,t')$ depend on the Green's function, and we maintain self-consistency using fixed-point iteration terminating when the maximum difference of the Green's function at subsequent iterates is less than $10^{-10}$.

We use the fifth-order backward differentiation formula multistep method to discretize the KBE. Gregory quadrature is used to discretize memory integrals to fifth-order accuracy. The discrete Lehmann representation \cite{KayeDLR}, implemented in \texttt{libdlr} \cite{kaye22_libdlr}, is used to discretize the imaginary time variables appearing in the vertical leg of the Keldysh contour, as in Ref.~\cite{kaye23_eqdyson}. We have confirmed that the compression of the Green's function using the HODLR scheme preserves conservation laws, namely the density and energy, to within the SVD truncation tolerance used in the scheme, which we set to $10^{-8}$.

In the pump-probe experiments, we measure the current, given by 
\begin{align}
    j(t) &= \text{Im}(\text{Tr}[\sigma_z(\Gamma_L(t)-\Gamma_R(t))])\\
    \Gamma_{R/L}(t) &= -i [G \ast \Delta_{L/R}]^<(t,t),
\end{align}
with $\ast$ the convolution operator on the three-legged Keldysh contour.

\section{Hartree-Fock}
\begin{figure}[h!]
    \centering
    \includegraphics[width=0.4\textwidth]{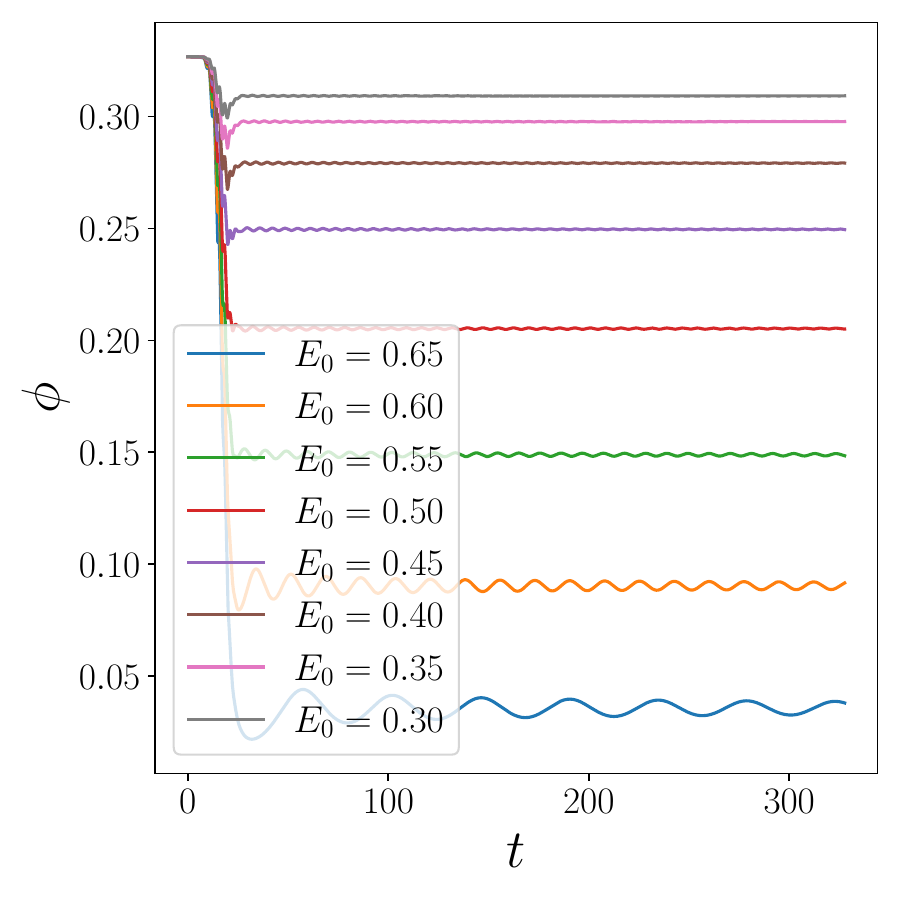}
    \caption{Time evolution of the order parameter $\phi$ within the Hartree-Fock approximation for increasing strength of the external pulse $E_0$.}
    \label{fig:HFOP}
\end{figure}

Solving the KBE allows for inclusion of dynamical fluctuations absent in a Hartree-Fock treatment of superconductors.  We show that including these correlations leads to qualitatively different dynamics, demonstrating the necessity of calculations beyond the mean field level.  In the main text, we observed chirping of the amplitude mode after the photo-excitation and we contrast these dynamics with the Hartree-Fock approximation, see Fig.~\ref{fig:HFOP}.  Already in equilibrium, the critical inverse temperature for the transition is $\beta_c=2$ which is much higher than in the 2nd Born approximation.

\begin{figure}[h!]
    \centering
    \includegraphics[width=0.5\textwidth]{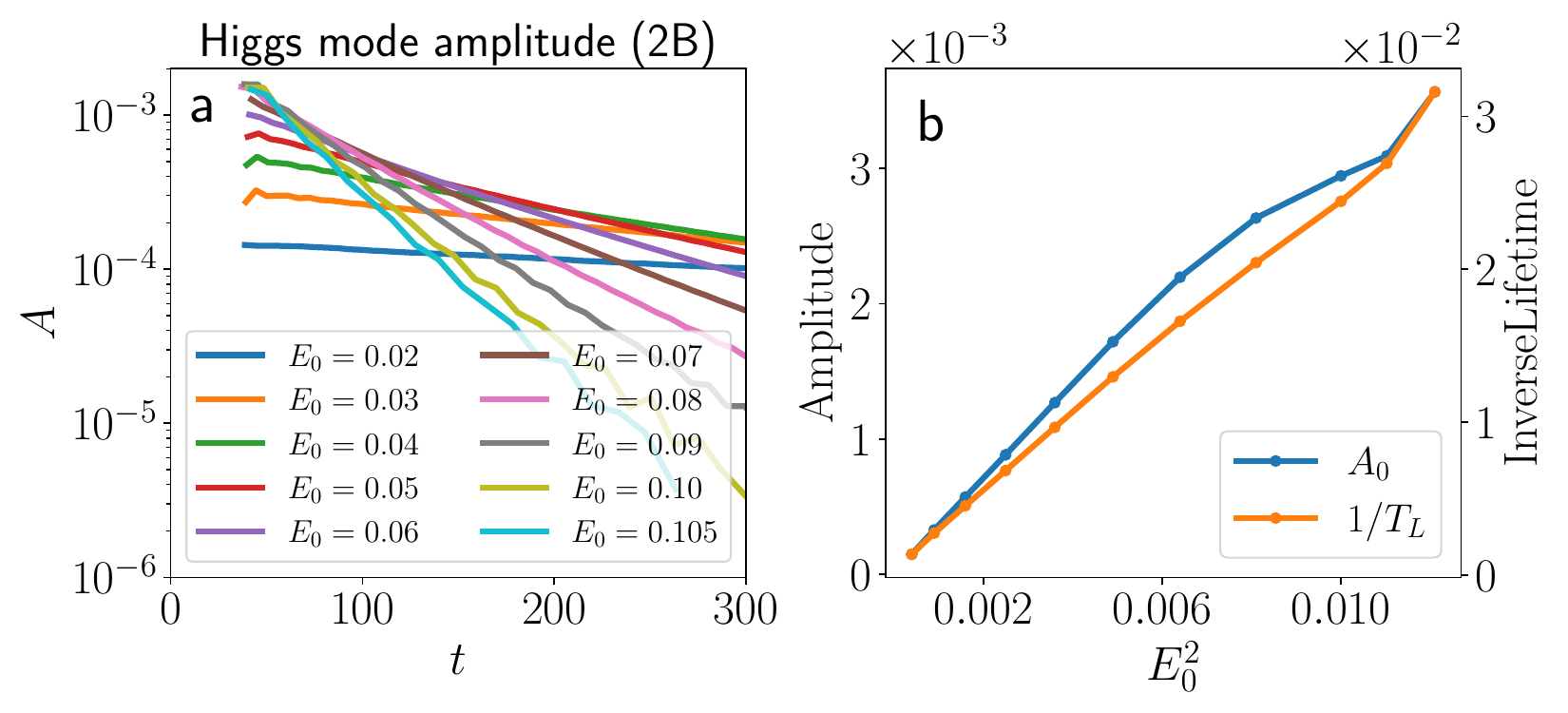}
    \newline\includegraphics[width=0.25\textwidth]{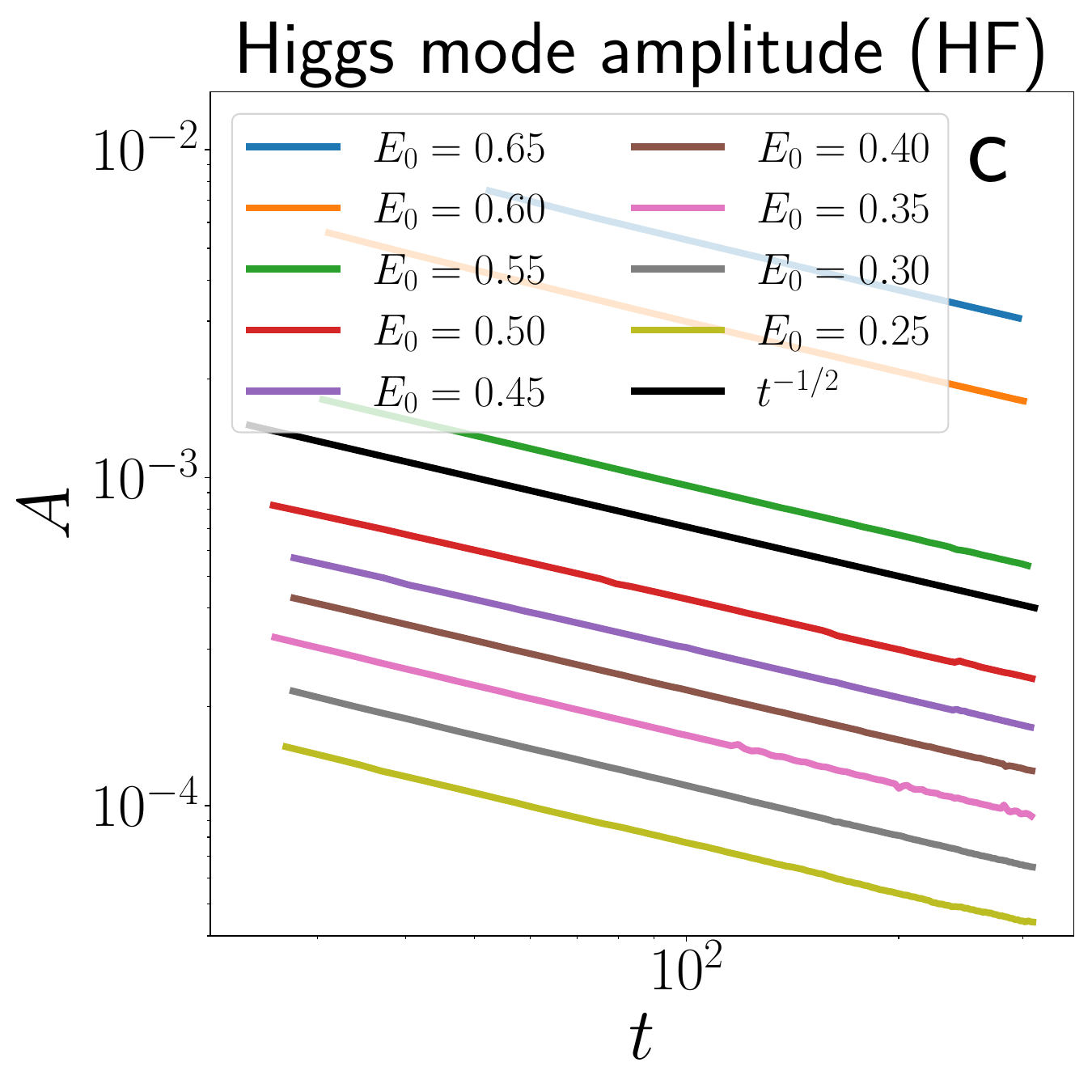}
    \caption{(a) Decay of the oscillating amplitude of the order parameter, $A$, for different excitation strengths $E_0$ within the 2nd Born ~(a) and the Hartree-Fock approximation~(c). The amplitude of the Hartree-Fock results is compared with the power-law decay $t^{-1/2}$. (b) The initial amplitude, $A_0$, and inverse lifetime, $1/T_L$, of the Higgs mode with increasing pump intensity for the 2B results.}
    \label{fig:envelope}
\end{figure}

We can use this order parameter data to confirm one of the main predictions of a Hartree-Fock treatment, which is a power-law decay of the Higgs mode~\cite{volkov73}. The amplitude of the Higgs oscillation in the Hartree-Fock and second Born approximation is presented in Fig.~\ref{fig:envelope}.  In the Hartree-Fock case, the expected power-law decay of $\phi\propto t^{-1/2}$ is shown to hold true over two orders of magnitude. However, when fluctuations are introduced through the 2nd Born self-energy, the Higgs mode instead decays exponentially~\cite{tsuji2015}, with the lifetime strongly depending on the excitation strength. 

\begin{figure}[h!]
    \centering
    \includegraphics[width=0.5\textwidth]{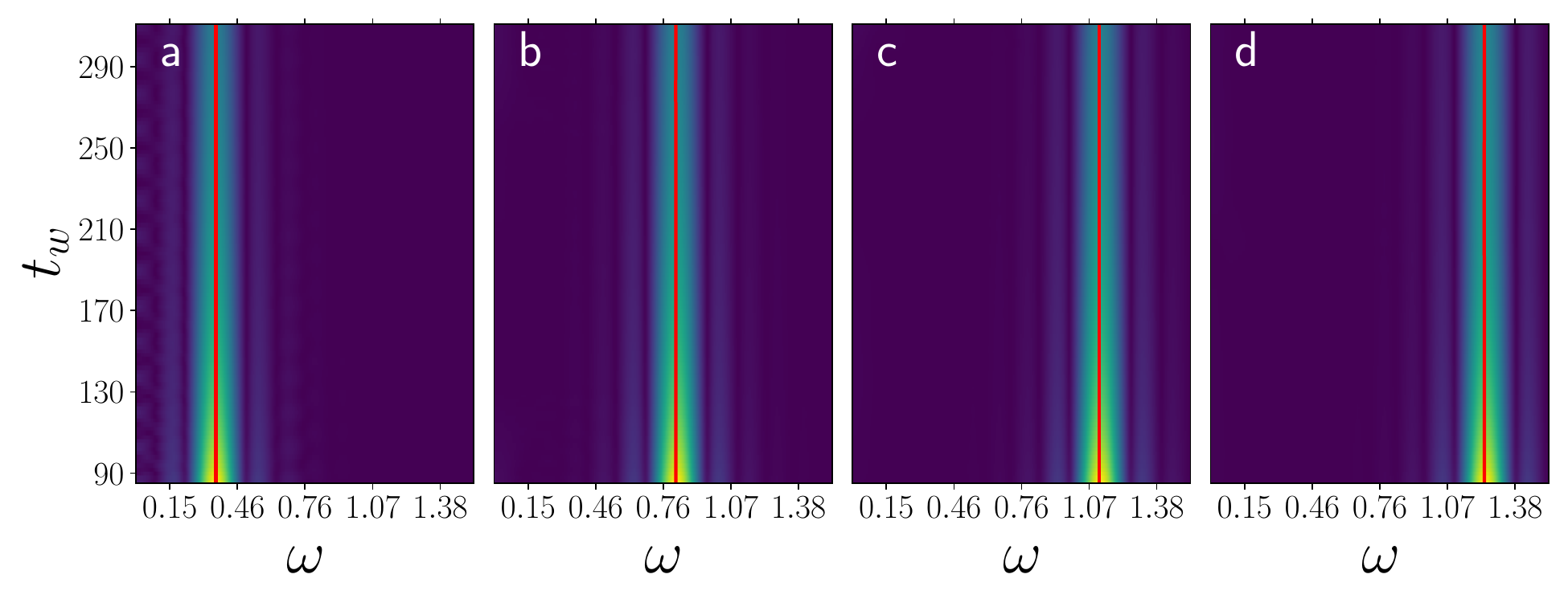}
    \caption{Windowed Fourier transform $F[\phi](t_w)$ of the order parameter centered around $t_w$ with background subtracted from the Hartree-Fock dynamics. Results for pump amplitudes (a) $E_0=0.6$, (b) $E_0=0.5$, (c) $E_0=0.4$, (d) $E_0=0.3$. The red line tracks the maximum of the spectrum, and varies by at most one unit of the frequency resolution.}
    \label{fig:HFchirp}
\end{figure}

Fig.~\ref{fig:HFOP} shows that when the pump field is turned off, the order parameter oscillates around a fixed value, in stark contrast to the continual decay present in the second order theory.  This suggests an immediate alteration of the potential energy surface within the system, suggesting an absence of chirping.  This absence of chirping is confirmed in Fig.~\ref{fig:HFchirp}, where the time dependence of the Higgs oscillation frequency is presented.  The frequency is constant for all pump strengths, with only noise less than 1\% of the original frequency.

\section{Numerical ranks in HODLR scheme}
\begin{figure}
    \centering
    \includegraphics[width=0.45\textwidth]{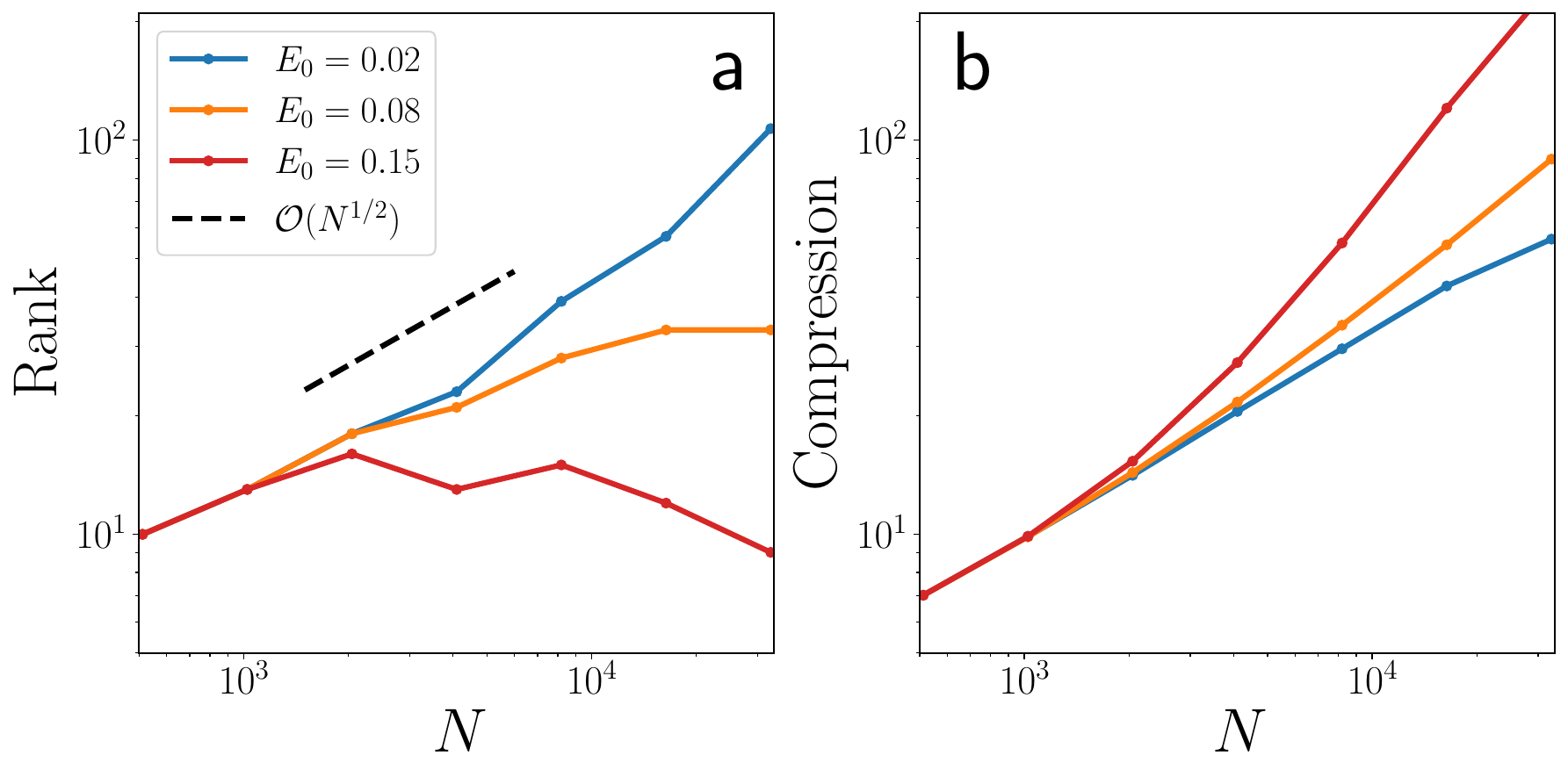}
    \caption{(a) Rank of the truncated SVD of the largest block in a HODLR decompostion of Green's function at timestep $T$.  Results for several different pump strengths are presented.  (b) Compression factor for Green's function and self-energy for the presented pump strengths.}
    \label{fig:rank}
\end{figure}

 Our time propagation scheme compresses the Green's function using a HODLR decomposition \cite{Kaye2021}. A truncated SVD is performed on each block of the decomposition, and updated on-the-fly, with singular values below $10^{-10}$ discarded. The numerical ranks of these blocks (within this precision) determine the compressibility of the Green's functions, and the performance of the scheme.

Fig.~\ref{fig:rank}(a) shows the growth of the maximum block rank $k$ with the number of time steps. At inverse temperature $\beta=6$, the system is in a disordered state, and the Green's functions decay exponentially on their off-diagonal. This leads to a rapid saturation of the ranks, giving an $\OO{N^2 \log N}$ computational complexity of the algorithm, and an $\OO{N \log N}$ memory complexity, with $N$ the number of time steps.
The behavior is similar for the simulations in which a low-temperature state is excited with a strong pump (i.e. $E_0=0.15$), destroying the superconducting state. 
At low temperature, in the ordered state, although the blocks are still numerically low rank with $k \ll N$, they grow as $k \propto \sqrt{N}$, leading to an $\OO{N^3 \log N}$ computational complexity and $\OO{N^{3/2} \log N}$ memory complexity. Nevertheless, while the memory usage, shown in Fig.~\ref{fig:rank}(b), reflects the expected scaling, the wall clock time shown in Fig.~\ref{fig:scaling} reflects approximately $\OO{N^2}$ scaling, suggesting that lower-scaling steps of the algorithm dominate throughout the time scale of this simulation.
\bibliography{mybib}
\end{document}